\documentclass[aps, prl, reprint, superscriptaddress]{revtex4-2}

\draft % marks overfull lines with a black rule on the right
\usepackage{graphicx}
\usepackage{dcolumn}
\usepackage{bm}
\usepackage{longtable}
\usepackage[colorlinks=true, allcolors=blue]{hyperref}
\usepackage{amsmath}

\begin{document}

%\preprint{}

\title{Nonlinear focusing of supercontinuum driven by intense mid-infrared pulses in gas-filled capillaries} %Title of paper



\author{Xiaohui Gao}
\email[]{gaoxh@utexas.edu}
%\thanks{}
\affiliation{Department of Physics, Shaoxing University, Shaoxing, Zhejiang 312000, China}

\date{\today}

\begin{abstract}
Strong mid-infrared light-matter interaction has attracted extensive attention as it opens up new frontiers in nonlinear optics. Here we observe through simulations a novel aspect of mid-infrared pulse dynamics in a high-pressure gas-filled capillary, where a pulse with a power well below the critical power for Kerr self-focusing undergoes an astonishing rise of the peak intensity following an extremely efficient spectral broadening.  This intensity enhancement is attributed to the Kerr-induced focusing of the supercontinuum.
Our study provides an interesting perspective for controlling the laser intensity with possible applications in nonlinear light conversion driven by mid-infrared pulses. 
\end{abstract}

\maketitle %\maketitle must follow title, authors, abstract

%\section{}
%\label{}
%\subsection{}
%\subsubsection{}
Advances in ultrafast laser technology have permitted the generation of high-energy mid-infrared pulses with remarkably short pulse duration. This opens up new regimes of strong-field light-matter interactions and provides an unparalleled opportunity to observe novel effects such as carrier wave shock~\cite{Whalen2014PRA, Hofstrand2020PRL} and to develop exciting applications spanning from the generation of remarkably broadband radiation~\cite{Mitrofanov2015SR, Mitrofanov2020O} to the generation of extremely high-energy harmonics~\cite{Popmintchev2012S}. For efficient nonlinear light conversion, it is essential to optimize the pulse propagation dynamics, as a stronger intensity leads to a higher microscopic yield, and the evolution of plasma density and intensity control the macroscopic yield through phase-matching conditions~\cite{Sun2017O}. 

Gas-filled capillaries have proven to be an unmatched system to explore the complex and rich pulse dynamics~\cite{Russell2014NP}. 
In monomode propagation where only the temporal evolution is involved, the anomalous group velocity dispersion enables efficient pulse self-compression~\cite{Balciunas2015NC} and phase-matched harmonic generation~\cite{Rundquist1998S}. At a higher intensity, the interference of multimodes excited by the self-focusing and ionization may be exploited for quasi-phase-matching~\cite{Zepf2007PRL} and spatial-temporal compression~\cite{Anderson2014PRA, Gao2018OL}. The intricate interaction of multiple spatial modes also leads to the observation of multi-dimensional optical soliton states~\cite{Safaei2020NP} and intermodal four-wave mixing~\cite{Piccoli2021NP}. For mid-infrared sources, capillary guiding has an advantage over the Kerr-induced filamentary propagation as it does not require a power exceeding the critical power for catastrophic self-focusing $P_\text{cr}$, which is extremely high for mid-infrared lasers as it increases quadratically with the wavelength $\lambda$. On the other hand, mid-infrared pulses propagating in capillaries suffer strong intermodal group velocity dispersion~\cite{Nagar2021OE} and high modal loss, leaving the pulse compression and intensity enhancement inefficient. 

In this paper, we numerically investigate the propagation dynamics of intense few-cycle mid-infrared laser pulses in a high-pressure gas-filled capillary. In contrast to previous study that the peak intensity decreases with the wavelength~\cite{Gao2018OL}, we find that a mid-infrared pulse with a power well below $P_\text{cr}$ experiences an unexpected rise of the peak intensity, and that the high intensity persists for a considerable length. This interesting behavior is caused by the nonlinear wavefront curvature induced by Kerr effect at the trailing edge of the pulse, which manifests itself as focusing of the supercontinuum.  

Simulations are performed using the carrier-resolved unidirectional pulse propagation equation~\cite{Kolesik2004PRE, Couairon2011EPJST, Brown2019}.
When a linearly polarized pulse with a radial symmetry is launched into a capillary of radius $a$, the electric field in terms of leaky modes is approximated as
\begin{equation}
E(r,t)=\sum_{n=1}^{n_\text{max}}E_n(t)J_0(k_{\perp,n}r),
\end{equation}
where $E_n(t)$ is the carrier-resolved complex electric field, $J_0(k_{\perp,n}r)$ is the zeroth-order Bessel function, $k_{\perp,n}=u_n/a$ is the axial wave number, $u_n$ is the $n$th zeros of $J_0$, $n$ is the mode index, and $r$ is the radial position. The series is truncated with $n_\text{max}$ for a practical reason. In the spectral domain, each modal field satisfies the following equation:
\begin{equation}
\frac{\partial E_n}{\partial z}=i[k_{z,n}(\omega)+i\alpha_n]E_n+\frac{i}{2k_{z,n}(\omega)}\frac{\omega^2}{\epsilon_0c^2}\left({\mathcal{P}_n}+i \frac{\mathcal{J}_n}{\omega}\right).
\label{eq2}
\end{equation}
Here $z$ is the propagation distance, $k_{z,n}$ is the longitudinal wave number
$
\sqrt{k^2(\omega)-k^2_{\perp,n}},
$
$k(\omega)=\omega n(\omega)/c$ is the total wave number, $\omega$ is angular frequency, $n(\omega)$ is the refractive index, $c$ is the speed of light in vacuum, $\alpha_n$ is the loss coefficient, $\epsilon_0$ is the vacuum permittivity, $\mathcal{P}$ is the nonlinear polarization, and $\mathcal{J}=\mathcal{J}_p+\mathcal{J}_a$ is the nonlinear polarization current which includes the nonlinear current due to plasma generation $\mathcal{J}_p$, and the nonlinear current due to ionization absorption $\mathcal{J}_\text{a}$. For $a\gg u_n\lambda/2\pi$, the loss coefficient is found to be~\cite{Marcatili1964TBSTJ},
\begin{equation}
\alpha_n=\frac{1}{2}\left(\frac{u_n}{2\pi}\right)^2\frac{\lambda^2}{a^3}\frac{1+\epsilon_\text{c}}{\sqrt{\epsilon_\text{c}-1}},
\end{equation}
where $\epsilon_\text{c}$ is the dielectric constant of the cladding. Nonlinear polarization and nonlinear current satisfy the following equations in the time domain , 
\begin{equation}
\mathcal{P}(t)=\epsilon_0\chi^{(3)}E(t)^3,
\end{equation}
\begin{equation}
\frac{\partial \mathcal{J}_p(t)}{\partial t}+\frac{\mathcal{J}_p(t)}{\tau_c}=\frac{e^2}{m_e}\rho_e(t) E(t),
\end{equation}
\begin{equation}
\mathcal{J}_a=\epsilon_0cn_0(\rho_0-\rho_e(t))\frac{W(|E(t)|)}{|E|^2}U_iE(t),
\end{equation}
where $\chi^{(3)}=4\epsilon_0cn_2n_0^2/3$ is the susceptibility, $n_2$ is the nonlinear refractive index, $n_0$ is the linear index, $\tau_c$ is the electron-neutral collision time, $e$ is the elementary charge, $m_e$ is the electron mass,  $\rho_0$ is the initial number density of the neutral gas, $\rho_e$ is the electron density, $W$ is the ionization rate calculated using the instantaneous field, $U_i$ is the ionization potential.  The electron density evolves with $t$ as
\begin{equation}
\frac{\partial \rho_e}{\partial t}=W(|E(t)|)(\rho_0-\rho_e).
\end{equation}
Here we have neglected avalanche ionization since it plays a minor role when the pulse duration is comparable to the collision time~\cite{Chimier2011PRB}. 

\begin{figure}[htbp]
\centering
\includegraphics[width=0.42\textwidth]{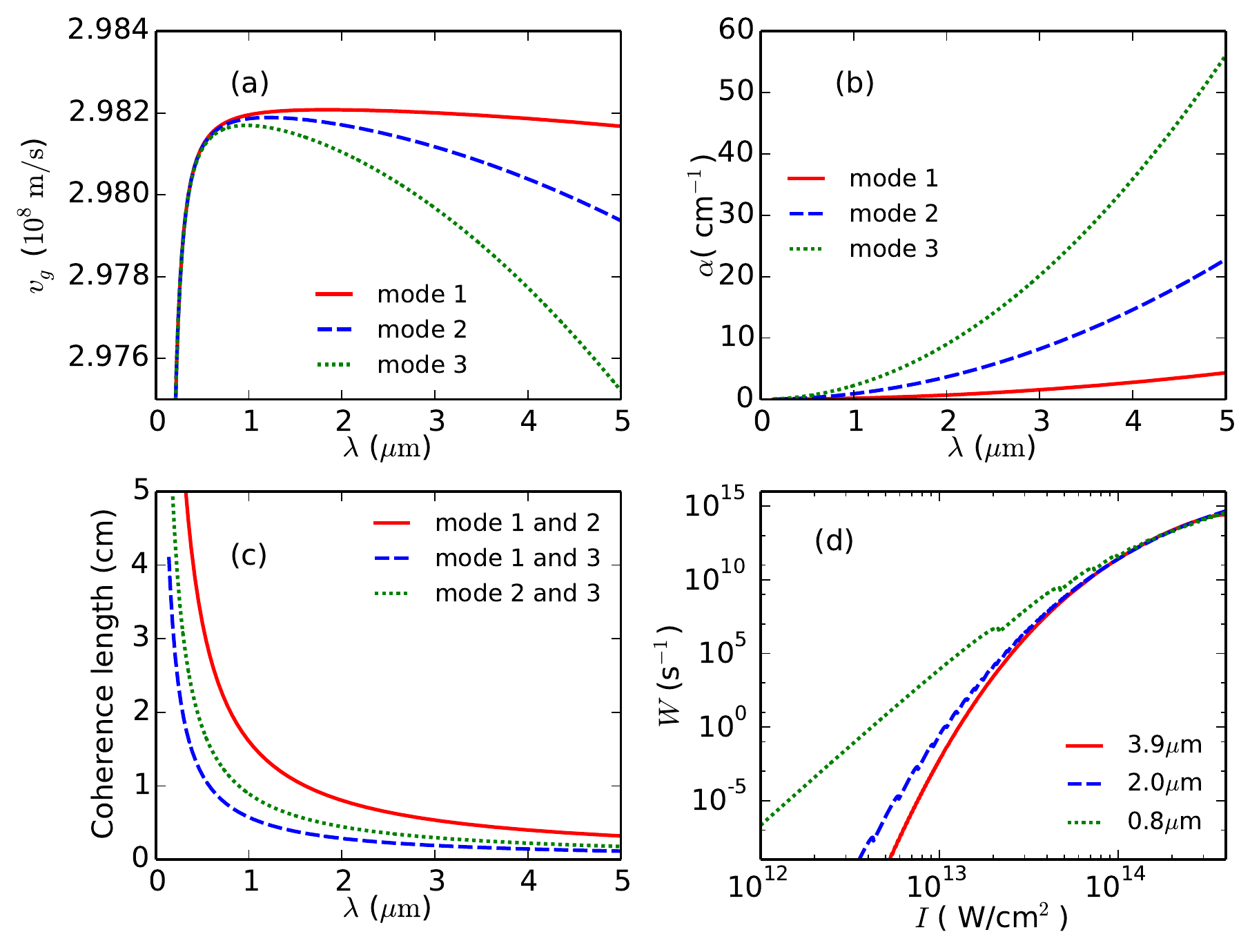}
\caption{(a)-(c) Variation of propagation properties with wavelength for a 100-$\mu$m radius capillary filled with a 20-bar argon gas: group velocity of the lowest three modes (a),  loss coefficient of the lowest three modes (b), and coherence length (c). (d) Ionization rate versus intensity for three wavelengths.}
\label{fig1} 
\end{figure}%
We consider a pulse with a Gaussian profile in temporal and spatial domain. The temporal envelope is $A(\tau)=\exp[-2\ln 2(\tau/\tau_p)^2]$, where $\tau_p$ is the FWHM pulse duration of the intensity envelope. The pulse is launched into the 100-$\mu$m radius capillary filled with a 20-bar argon gas.  The beam waist satisfies $w=0.645a$ so that a maximum energy is coupled into the fundamental mode. Values of $n_2/p =1.0\times10^{-19}\,$cm$^2$/W/bar~\cite{Zahedpour2015OL} and $\tau_c/p = 190\,$fs/bar with $p$ the pressure in unit of bar are used. The group velocity dispersion $v_g$, the loss coefficient $\alpha_n$ for the first three modes, and the coherence length between mode $n$ and $n'$ calculated using $l=\pi/(k_{z,n}-k_{z,n'})$ are plotted against the driving laser wavelength in Fig.~\ref{fig1}(a), Fig.~\ref{fig1}(b), and Fig.~\ref{fig1}(c), respectively, which exhibit a strong modal dispersion and large loss coefficient at mid-infrared wavelength. An ionization rate $W$ valid for an arbitrary frequency is used~\cite{Popruzhenko2008PRL}, and its intensity dependence is plotted in Fig.~\ref{fig1}(d) for three wavelengths.  A value of $n_\text{max}=20$ is used. The convergence of the results is checked by comparing output using different $n_\text{max}$. The active wavelength window is from 100-nm to 20-$\mu$m.  

\begin{figure}[htbp]
\centering
\includegraphics[width=0.38\textwidth]{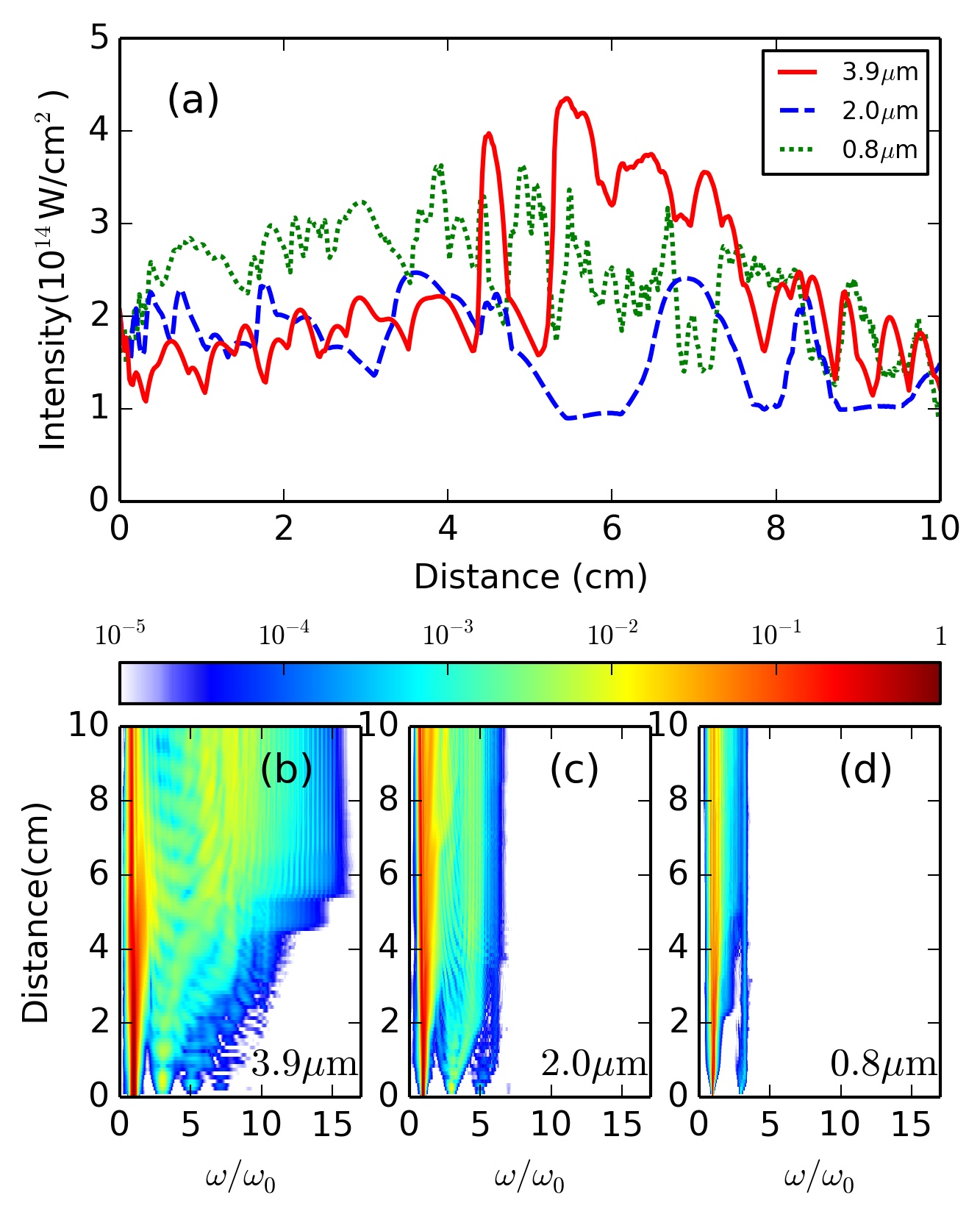}
\caption{(a) Evolution of the peak intensity over a 10-cm propagation distance for different driving wavelengths: $\lambda$=$3.9\,\mu$m (solid red line), $\lambda$=$2.0\,\mu$m (dashed blue line), and $\lambda$=$0.8\,\mu$m (dotted green line). (b)-(d) Spectral evolution for the three cases: $\lambda$=$3.9\,\mu$m (b), $\lambda$=$2.0\,\mu$m (c), and $\lambda$=$0.8\,\mu$m (d).}
\label{fig2} 
\end{figure}%
Figure~\ref{fig2}(a) shows the evolution of peak intensity over a 10-cm propagation distance when a 180-$\mu$J, 25-fs laser pulse with different wavelengths is launched into the capillary. The input peak power $P_\text{in}$ of 6.76\,GW corresponds to 0.60$P_\text{cr}$ at 3.9\,$\mu$m, 2.28$P_\text{cr}$ at 2\,$\mu$m, and 14.3$P_\text{cr}$ at 0.8\,$\mu$m, respectively. We observe an initial rise in the laser intensity for $\lambda=0.8\,\mu$m and 2.0\,$\mu$m due to self-focusing. The higher peak intensity at 0.8\,$\mu$m is consistent with previous study that the spatiotemporal localization is favored at a shorter wavelength~\cite{Gao2018OL}. The intensity evolution using 3.9-$\mu$m driving laser wavelength shows a gentle increase over the first 4-cm, with a modulation period of 0.8-cm. However, the dynamics is dramatically altered for $z>4$ cm. 
A sudden rise appears in the intensity with the same modulation period of approximately 0.8 cm, reaching a maximum intensity significantly higher than that using near-infrared laser pulse. The intensity is maintained above $3\times10^{14}$\,W/cm$^2$ for approximately 2 cm. 
Figure~\ref{fig2}(b)-(d) show the spectral evolution of the three cases with a spectral intensity range of 50\,dB. For $\lambda=3.9\,\mu$m, we observe an extremely broad supercontinuum extending beyond the seventh harmonics at $z=4\,$cm, while only third harmonics of 0.8\,$\mu$m are observed in Fig.~\ref{fig2}(d). Note the absence of higher harmonics is not due to spectral filtering in the simulation. Efficient harmonic generation is a distinct feature of mid-infrared propagation~\cite{Mitrofanov2015SR, Mitrofanov2020O} and is an essential ingredient in novel effects such as radiation-arrested pulse collapse ~\cite{Panagiotopoulos2015NP} and efficient terahertz generation~\cite{Koulouklidis2020NC}. It may also play an important role in our unexpected behavior of intensity rise.

\begin{figure}[htbp]
\centering
\includegraphics[width=0.45\textwidth]{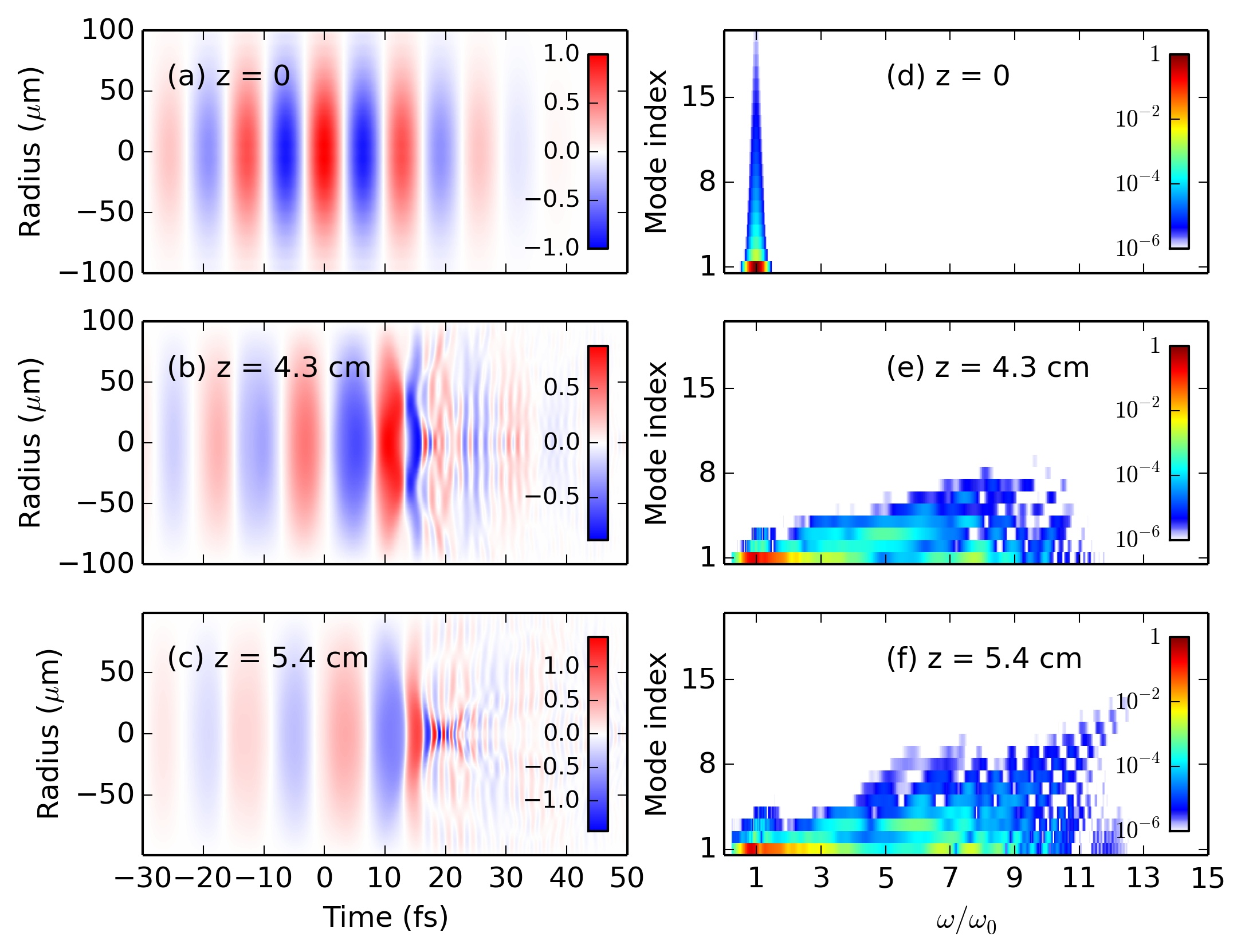}
\caption{Spatiotemporal profiles of the electric fields at (a) $z=0$, (b) $z=4.3\,$cm, (c) $z=5.4\,$cm. Modal-resolved spectral intensity at (d) $z=0$, (e) $z=4.3\,$cm, (f) $z=5.4\,$cm. }
\label{fig3} 
\end{figure}%
While a gradual increase in intensity is usually caused by pulse compression, a sudden rise of laser intensity is often a manifestation of catastrophic focusing. To elucidate the origin of the intensity enhancement, we present the spatio-temporal field and the spectral content of each modes at $z=0$, 4.3\,cm, and 5.4\,cm in Fig.~\ref{fig3}(a)-(c) and Fig.~\ref{fig3}(d)-(e), respectively. The spatiotemporal field and modal-resolved spectral intensity of an unchirped Gaussian pulse with zero carrier envelope phase are shown in Fig.~\ref{fig3}(a) and Fig.~\ref{fig3}(d), respectively. As the pulse propagates, the leading pulse is red-shifted and the trailing pulse is strongly blue-shifted, and the spectrum is remarkably broadened due to self-phase modulation, harmonic generation, and cross-phase modulation.  Figure~\ref{fig3}(b) clearly shows a nonlinear curvature of the wavefront at the trailing edge of the pulse. The resulting focusing effect produces a spatio-temporal localized hotspot with a maximum field of approximately 1.4 times the initial field, as shown in Fig.~\ref{fig3}(c). The mode-frequency profile of the spectral energy density at $z=5.4$\,cm shown in Fig.~\ref{fig3}(f) confirms that the hotspot is due to the higher order modes at higher frequencies. High order modes are crucial to describe the observed dynamics of sudden increase of intensity during the propagation. The wavelength is short for these high order modes so the condition $a\gg u_n\lambda/2\pi$ remains satisfied.

\begin{figure}[ht!]
\centering
\includegraphics[width=0.4\textwidth]{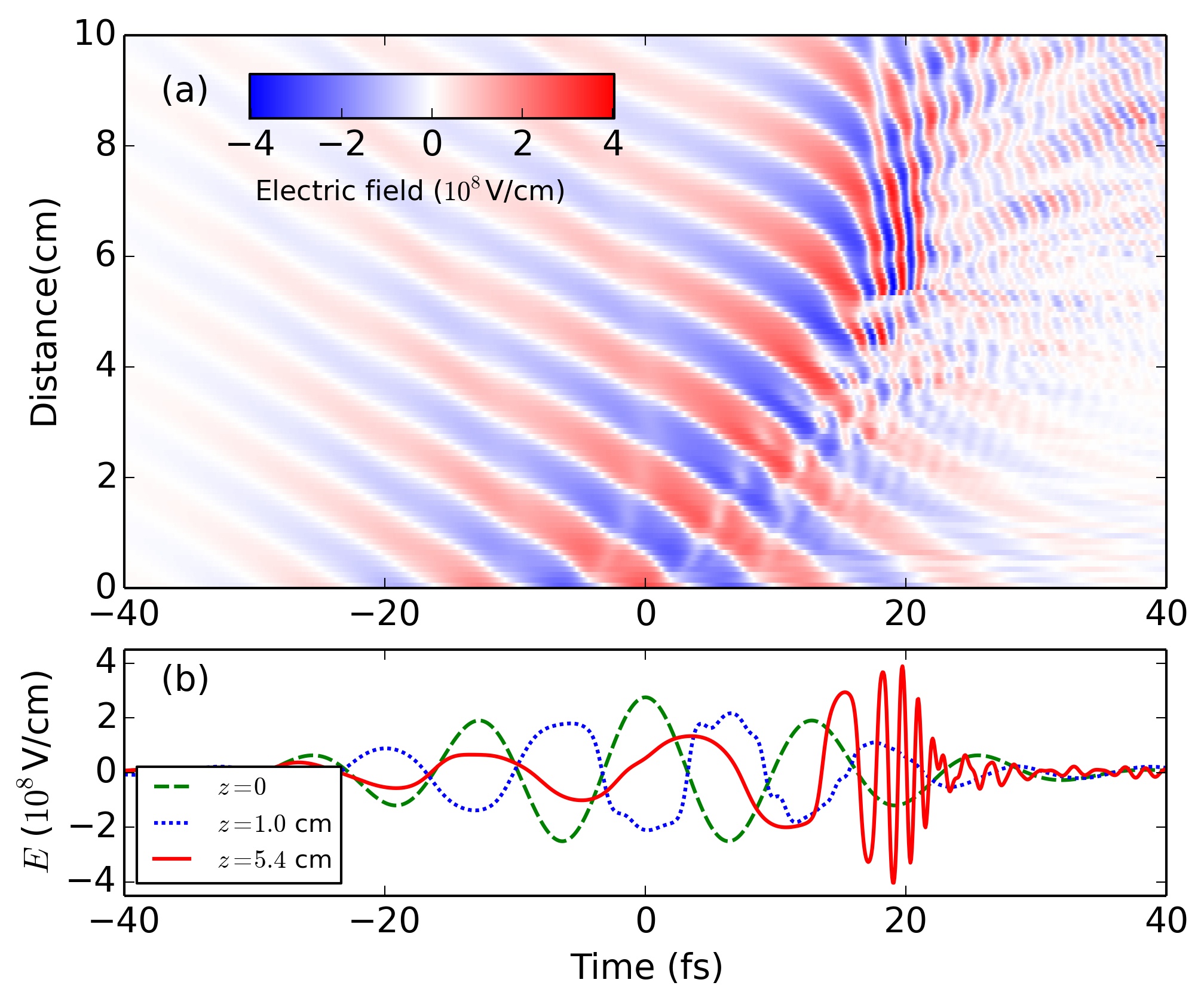}
\caption{(a) Evolution of the on-axis temporal field $E(t)$ over a 10-cm distance. (b) Plots of $E(t)$ at $z=0$ (dashed green line), $z=1.0\,$\,cm (dotted blue line), and $z=5.4\,$cm (solid red line).}
\label{fig4} 
\end{figure}%
Figure~\ref{fig4}(a) shows the on-axis temporal evolution of laser field of $\lambda$=3.9\,$\mu$m over a 10-cm distance, and Fig.~\ref{fig4}(b) shows the variation of axial field with the time at three locations: $z=0$, 1.0\,cm, and 5.4\,cm. The phase of field at $z=1.0\,$ cm (dotted blue line in Fig.~\ref{fig4}(b)) is roughly shifted by $\pi$ with respect to that at $z=0$ (dashed green line in Fig.~\ref{fig4}(b)).  As is shown in Fig.~\ref{fig2}(a), the intensity at $z=1.0$\, cm reaches a local minimum. Thus the periodic oscillation of the instantaneous intensity at $\lambda=3.9$\,$\mu$m is due to the evolution of the carrier-envelope phase. In contrast, the pulse at $\lambda=0.8$\,$\mu$m consists of many optical cycles, and the carrier-envelope phase effect is washed out. The interference among different spatial modes causes intensity oscillation with various oscillation periods.
%Instead, the intensity oscillation is due to the interference of different spatial modes where the complicate period is due to the beating of multiple modes.

\begin{figure}[htbp!]
\centering
\includegraphics[width=0.4\textwidth]{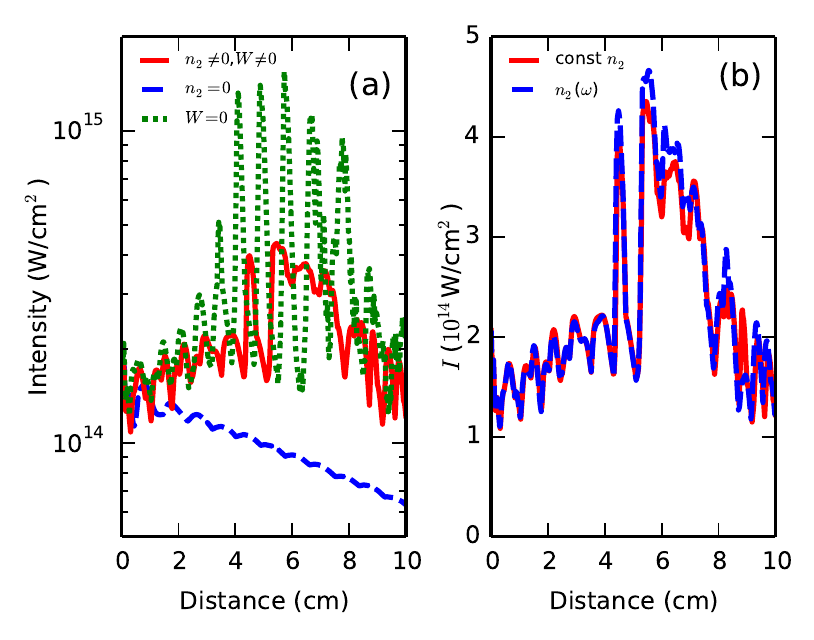}
\caption{(a) Propagation of 25-fs 3.9-$\mu$m pulse when the ionization is switched off (dotted green line) or $n_2$ is switched off (dashed blue line) (b) Propagation dynamics with a constant $n_2$ (solid red line) and frequency dependent $n_2(\omega)$ (dashed blue line).}
\label{fig5} 
\end{figure}%
The curvature of the wavefront in Fig.~\ref{fig3} indicates the on-axis refractive index is greater than that at the periphery. This can be caused by the Kerr effect or index variation of a plasma due to electron density distribution~\cite{Gao2019OLb} or spatial chirp~\cite{ Gao2020OL}. To clarify whether this is Kerr-induced focusing, we artificially switch off the nonlinear terms, and the results are presented in Fig.~\ref{fig5}(a). The dotted green curve shows peak intensity versus the distance when the ionization is switched off. The pronounced field enhancement confirms that this behavior is due to the Kerr effect. The dashed blue line shows the intensity evolution when $n_2=0$ and the ionization becomes the only source of nonlinearity. A minor Kerr-induced focusing does not require the collapse threshold~\cite{Marburger1975PQE}. Self-focusing with a power slightly below the critical power can also occur in capillaries due to self-compression~\cite{Crego2019SR}. The dramatic increase in the laser intensity in our simulations should be a catastrophic focusing and differs from these cases. As Fig.~\ref{fig3}(f) shows, higher order modes are populated mainly at high frequencies. At shorter wavelength, the focusing threshold is substantially reduced. This may bear similarity with  the nonlinear focusing in stimulated Raman scattering when the power falls below the critical power~\cite{Hafizi2015OL}.

Because of the extreme spectral broadening, the dispersion of $n_2$ may not be negligible. To investigate it, $\mathcal{P}_n(\omega)$ in Eq.~\ref{eq2} is replaced with $n_2(\omega)/n_2(\omega_0)\mathcal{P}_n(\omega)$, and $n_2(\omega)=A/(1-B\omega^2)$ , where $A$ and $B$ are chosen to match $n_2$ at 200-nm and 3.9-$\mu$m~\cite{Bree2011PRL}. The active frequency is from 200-nm as there is a resonance of $n_2$ below 200-nm due to two photon absorption. 
Simulations shown in Fig.~\ref{fig5}(b) confirms the sudden rise of the intensity in both cases.  Indeed, the dispersion of $n_2$ increases the focused intensity slightly. This can be expected as $n_2$ gets higher at shorter wavelength. Since the difference is not substantial, our simulations using constant $n_2$ is valid. 

\begin{figure}[htbp]
\centering
\includegraphics[width=0.45\textwidth]{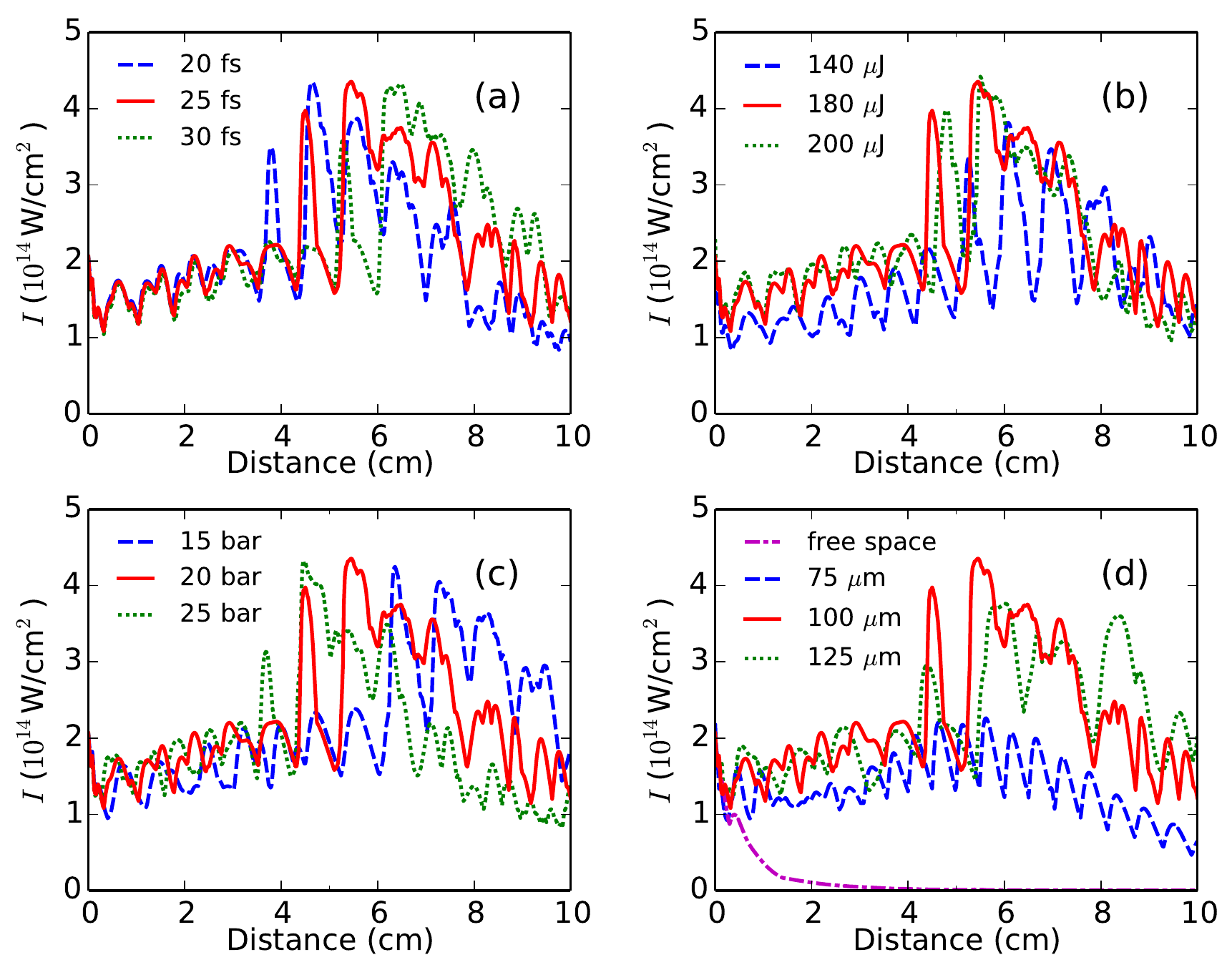}
\caption{Parameter dependencies of the peak intensity evolution over a 10-cm distance: (a) pulse duration (b) pulse energy (c) gas pressure (d) capillary size.}
\label{fig6} 
\end{figure}% 
Figure~\ref{fig6}(a) and \ref{fig6}(b) show the dependence of the pulse dynamics on the pulse duration and energy, respectively. For shorter pulse duration, the intensity rises earlier. This may be attributed to a broader initial spectrum from a shorter pulse. The peak intensity saturates with increasing energy. This is expected as the intensity will be also clamped by the plasma effect. Figure~\ref{fig6}(c) show the pressure dependence.  While the collapse distance decreases with the pressure, the peak intensity remains similar. Figure~\ref{fig6} (d) shows the capillary size dependence. In this simulation, we adjust the beam waist and input energy to maintain the same initial intensity. At $a=75\,\mu$m, we only observe a smooth increase of the intensity. The absence of abrupt rise in intensity could be high modal loss at a smaller size. At $a=125\,\mu$m, the enhancement is also weaker. This may be due to small group velocity dispersion as the size increases. Thus there exists a trade off between the loss and dispersion. Here the dash-dotted magenta line is the case for free space propagation, where the intensity drops quickly as the power is not sufficient to counteract the diffraction. 

In conclusion, we demonstrate a novel nonlinear focusing phenomenon when an intense few-cycle mid-infrared pulses propagates in high-pressure gas-filled capillaries. Despite of $P<P_\text{cr}$, the nonlinear phase acquired by the extremely blue-shifted supercontinuum at the trailing edge of the pulse induces strong wavefront curvature, leading to a catastrophic focusing with a high intensity persisting for several centimeters. This elongated interaction length at high intensity and high atomic density of high-pressure gases are desirable for nonlinear light conversion.  Our results add to the rich dynamics of mid-infrared laser pulses and could be relevant to high-field applications such as high-order harmonic generation.

\noindent{\bf{Funding.}} Natural Science Foundation of Zhejiang Province (LY19A040005).

% If in two-column mode, this environment will change to single-column format so that long equations can be displayed. 
%\begin{widetext}
%$$\mbox{put long equation here}$$
%\end{widetext}

% Figures should be put into the text as floats. 
% Use the graphics or graphicx packages (distributed with LaTeX2e).
% Put the label that you will use with \ref{} command in the braces of the \label{} command.
%
% \begin{figure}
% \end{figure}

% \begin{table}
% \end{tabular}
% \end{table}

%\begin{acknowledgments}
% Put your acknowledgments here.
%\end{acknowledgments}

\bibliographystyle{aipnum4-2}

\begin{thebibliography}{10}
\newcommand{\enquote}[1]{``#1''}

\bibitem{Whalen2014PRA}
P.~Whalen, P.~Panagiotopoulos, M.~Kolesik, and J.~V. Moloney,
  {\protect{Phys. Rev. A}} \textbf{89}, 023850 (2014).

\bibitem{Hofstrand2020PRL}
A.~Hofstrand and J.~V. Moloney,  {\protect{Phys. Rev. Lett.}}
  \textbf{124}, 043901 (2020).

\bibitem{Mitrofanov2015SR}
A.~V. Mitrofanov, A.~A. Voronin, D.~A. Sidorov-Biryukov, A.~Pug{\v{z}}lys,
  E.~A. Stepanov, G.~Andriukaitis, T.~Fl{\"o}ry, S.~Ali{\v{s}}auskas, A.~B.
  Fedotov, A.~Baltu{\v{s}}ka, and A.~M. Zheltikov,  {\protect{Sci.
  Rep.}} \textbf{5}, 8368 (2015).

\bibitem{Mitrofanov2020O}
A.~V. Mitrofanov, D.~A. Sidorov-Biryukov, M.~M. Nazarov, A.~A. Voronin, M.~V.
  Rozhko, A.~D. Shutov, S.~V. Ryabchuk, E.~E. Serebryannikov, A.~B. Fedotov,
  and A.~M. Zheltikov,  {\protect{Optica}} \textbf{7}, 15 (2020).

\bibitem{Popmintchev2012S}
T.~Popmintchev, M.~C. Chen, D.~Popmintchev, P.~Arpin, S.~Brown, S.~Alisauskas,
  G.~Andriukaitis, T.~Balciunas, O.~D. Mucke, A.~Pugzlys, A.~Baltuska, B.~Shim,
  S.~E. Schrauth, A.~Gaeta, C.~Hernandez-Garcia, L.~Plaja, A.~Becker,
  A.~Jaron-Becker, M.~M. Murnane, and H.~C. Kapteyn,
  {\protect{Science}} \textbf{336}, 1287 (2012).

\bibitem{Sun2017O}
H.-W. Sun, P.-C. Huang, Y.-H. Tzeng, J.-T. Huang, C.~D. Lin, C.~Jin, and M.-C.
  Chen,  {\protect{Optica}} \textbf{4}, 976 (2017).

\bibitem{Russell2014NP}
P.~S.~J. Russell, P.~Hölzer, W.~Chang, A.~Abdolvand, and J.~C. Travers,
  {\protect{Nat. Photonics}} \textbf{8}, 278 (2014).

\bibitem{Balciunas2015NC}
T.~Balciunas, C.~Fourcade-Dutin, G.~Fan, T.~Witting, A.~A. Voronin, A.~M.
  Zheltikov, F.~Gerome, G.~G. Paulus, A.~Baltuska, and F.~Benabid,
  {\protect{Nat. Commun.}} \textbf{6}, 6117 (2015).

\bibitem{Rundquist1998S}
A.~Rundquist, C.~G. Durfee~III, Z.~Chang, C.~Herne, S.~Backus, M.~M. Murnane,
  and H.~C. Kapteyn, {\protect{Science}} \textbf{280}, 1412
  (1998).

\bibitem{Zepf2007PRL}
M.~Zepf, B.~Dromey, M.~Landreman, P.~Foster, and S.~Hooker,
  {\protect{Phys. Rev. Lett.}} \textbf{99}, 143901 (2007).

\bibitem{Anderson2014PRA}
P.~N. Anderson, P.~Horak, J.~G. Frey, and W.~S. Brocklesby,
  {\protect{Phys. Rev. A}} \textbf{89}, 013819 (2014).

\bibitem{Gao2018OL}
X.~Gao, G.~Patwardhan, B.~Shim, T.~Popmintchev, H.~C. Kapteyn, M.~M. Murnane,
  and A.~L. Gaeta, {\protect{Opt. Lett.}} \textbf{43}, 3112
  (2018).

\bibitem{Safaei2020NP}
R.~Safaei, G.~Fan, O.~Kwon, K.~Légaré, P.~Lassonde, B.~E. Schmidt,
  H.~Ibrahim, and F.~Légaré, {\protect{Nat. Photonics}}
  \textbf{14}, 733 (2020).

\bibitem{Piccoli2021NP}
R.~Piccoli, J.~M. Brown, Y.-G. Jeong, A.~Rovere, L.~Zanotto, M.~B. Gaarde,
  F.~Légaré, A.~Couairon, J.~C. Travers, R.~Morandotti, B.~E. Schmidt, and
  L.~Razzari, {\protect{Nat. Photonics}} \textbf{15}, 884 (2021).

\bibitem{Nagar2021OE}
G.~C. Nagar and B.~Shim, {\protect{Opt. Express}} \textbf{29},
  27416 (2021).

\bibitem{Kolesik2004PRE}
M.~Kolesik and J.~Moloney, {\protect{Phys. Rev. E}} \textbf{70},
  036604 (2004).

\bibitem{Couairon2011EPJST}
A.~Couairon, E.~Brambilla, T.~Corti, D.~Majus, O.~d.~J.
  Ram{\'\i}rez-G{\'o}ngora, and M.~Kolesik, {\protect{Eur. Phys.
  J. Spec. Top.}} \textbf{199}, 5 (2011).

\bibitem{Brown2019}
J.~M. Brown, \enquote{laser-propagation,}
  \url{https://github.com/brownjm/laser-propagation} (2019). Accessed:
  2022-06-06.

\bibitem{Marcatili1964TBSTJ}
E.~Marcatili and R.~Schmeltzer, {\protect{Bell Syst. Tech. J.}}
  \textbf{43}, 1783 (1964).

\bibitem{Chimier2011PRB}
B.~Chimier, O.~Ut\'{e}za, N.~Sanner, M.~Sentis, T.~Itina, P.~Lassonde,
  F.~L\'{e}gar\'{e}, F.~Vidal, and J.~C. Kieffer, {\protect{Phys.
  Rev. B}} \textbf{84}, 094104 (2011).

\bibitem{Zahedpour2015OL}
S.~Zahedpour, J.~K. Wahlstrand, and H.~M. Milchberg,
  {\protect{Opt. Lett.}} \textbf{40}, 5794 (2015).

\bibitem{Popruzhenko2008PRL}
S.~V. Popruzhenko, V.~D. Mur, V.~S. Popov, and D.~Bauer,
  {\protect{Phys. Rev. Lett.}} \textbf{101}, 193003 (2008).

\bibitem{Panagiotopoulos2015NP}
P.~Panagiotopoulos, P.~Whalen, M.~Kolesik, and J.~V. Moloney,
  {\protect{Nat. Photonics}} \textbf{9}, 543 (2015).

\bibitem{Koulouklidis2020NC}
A.~D. Koulouklidis, C.~Gollner, V.~Shumakova, V.~Y. Fedorov, A.~Pug\u{z}lys,
  A.~Baltu\u{s}ka, and S.~Tzortzakis, {\protect{Nat. Commun.}}
  \textbf{11}, 292 (2020).

\bibitem{Gao2019OLb}
X.~Gao, G.~Patwardhan, B.~Shim, and A.~L. Gaeta, {\protect{Opt.
  Lett.}} \textbf{44}, 5888 (2019).

\bibitem{Gao2020OL}
X.~Gao and B.~Shim, {\protect{Opt. Lett.}} \textbf{45}, 6434
  (2020).

\bibitem{Marburger1975PQE}
J.~H. Marburger, {\protect{Prog. Quantum Electron.}} \textbf{4},
  35 (1975).

\bibitem{Crego2019SR}
A.~Crego, E.~Conejero~Jarque, and J.~San~Roman, {\protect{Sci.
  Rep.}} \textbf{9}, 9546 (2019).

\bibitem{Hafizi2015OL}
B.~Hafizi, J.~P. Palastro, J.~R. Peñano, D.~F. Gordon, T.~G. Jones, M.~H.
  Helle, and D.~Kaganovich, {\protect{Opt. Lett.}} \textbf{40},
  1556 (2015).

\bibitem{Bree2011PRL}
C.~Br\'ee, A.~Demircan, and G.~Steinmeyer, {\protect{Phys. Rev.
  Lett.}} \textbf{106}, 183902 (2011).

\end{thebibliography}
%aipnum4-2.bst 2019-01-14 (MD) hand-edited version of apsrev4-1.bst

\end{document}